\documentclass[twocolumn,           
               showpacs,            
               preprintnumbers,     
               aps,                 
               prd,          	    
               a4paper,             
               superscriptaddress,  
               unsortedaddress,     
               nofootinbib,         
               tightenlines,        
               floats               
               ]{revtex4}

\usepackage{latexsym}
\usepackage{amsmath,amssymb}        
\usepackage[draft=false]{hyperref}

\begin{document}


\title{Primordial black holes in braneworld cosmologies: Accretion after 
formation}

\author{Raf Guedens}
\affiliation{DAMTP, Centre for Mathematical Sciences,
 Cambridge University, Wilberforce Road, 
            Cambridge CB3 0WA, United 
Kingdom}
\author{Dominic Clancy}
\affiliation{Astronomy Centre, University of Sussex, 
             Brighton BN1 9QJ, United 
Kingdom}
\author{Andrew R.~Liddle}
\affiliation{Astronomy Centre, University of Sussex, 
             Brighton BN1 9QJ, United 
Kingdom}
\date{\today} 
\pacs{98.80.Cq \hfill astro-ph/0208299}
\preprint{astro-ph/0208299}

\begin{abstract}
We recently studied the formation and evaporation of primordial black holes in a 
simple braneworld cosmology, namely Randall--Sundrum Type II. Here we study the 
effect of accretion from the cosmological background onto the black holes after 
formation. While it is generally believed that in the standard cosmology such 
accretion is of negligible importance, we find that during the high-energy 
regime of braneworld cosmology accretion can be the dominant effect and lead to 
a mass increase of potentially orders of magnitude. However, unfortunately the 
growth is exponentially sensitive to the accretion efficiency, which cannot be 
determined accurately. Since accretion becomes unimportant once the high-energy 
regime is over, it does not affect any constraints expressed at the time of 
black hole evaporation, but it can change the interpretation of those 
constraints in terms of early Universe formation rates.
\end{abstract}

\maketitle

\section{Introduction}

Primordial black holes (PBHs) are relics which may form in the early Universe 
\cite{Hawk71}
and survive to have effects in later epochs, thus shedding light on the physical 
processes operating during the Universe's early stages. In a recent paper 
(Guedens, Clancy and Liddle \cite{GCL}), we considered the physics of black 
holes in a popular variant to the standard cosmology, the Randall--Sundrum Type 
II braneworld cosmology \cite{RSII}. We showed that the presence of the fifth 
(AdS) 
dimension could significantly modify the properties of black holes, and indeed 
that black holes evaporating at key epochs of the Universe might behave as 
higher-dimensional objects.

The results of that paper assumed that after formation the dominant process 
affecting the mass of the black holes was Hawking evaporation. In the standard 
cosmology it is believed that the neglect of accretion of material from the 
cosmological background is a very good approximation, except perhaps immediately 
after formation, though the calculations are rather uncertain. However, we 
remarked that it was not necessarily true that accretion would also be 
negligible during the high-energy regime. The purpose of this paper is to 
investigate accretion during the high-energy regime.\footnote{As we were 
completing this paper, a paper appeared by Majumdar \cite{Maj}, using the 
formalism of our earlier paper \cite{GCL} to investigate this same question. Our 
results are in broad agreement, though we will highlight the key differences.} 
We will find that accretion can indeed be the dominant effect in the high-energy 
regime, though accurate calculation of its effects is presently impossible.

Our calculations follow the notation of Ref.~\cite{GCL}, and we will consider 
both the standard cosmology and the Randall--Sundrum Type II cosmology, which 
features a high-energy regime where the Friedmann equation is modified. The 
AdS radius of the extra dimension will be denoted $l$, and $M_4$ and $l_4$ 
indicate the usual four-dimensional Planck mass and length (we set $c=1$). The 
four-dimensional cosmological constant set to zero. We 
use $M$ for the 
black hole mass and $r_0$ for the Schwarzschild radius. If $r_0 < l$ then the 
black hole will be effectively five-dimensional and the appropriate 
five-dimensional Schwarzschild solution is used.

\section{Accretion formalism}

Calculations of PBH accretion in the standard cosmology have a long history but 
are plagued with significant uncertainties. First of all, while it is known that 
black holes forming during radiation domination cannot be much smaller than the 
horizon size (otherwise pressure forces would prevent collapse), it is unclear 
how much smaller they might be. Secondly, it is unclear precisely how efficient 
a black hole might be at absorbing incoming radiation from a cosmological 
background, or how important the effect of backreaction might be. Results prove 
extremely sensitive to assumptions made concerning these quantities. While early 
work by Zel'dovich and Novikov \cite{ZelNov} speculated that PBHs might even be 
able to grow as fast as 
the horizon, subsequent work, especially by Carr and Hawking 
\cite{CarrHawk,Carr75}, 
made a convincing case that such growth could not occur, and moreover that once 
the PBH became significantly smaller than the horizon accretion would become 
very inefficient. This view is now widely held, though there are 
papers where accretion is found to be important, including interesting work by 
Hacyan \cite{Hacyan} in which incoming 
radiation was described by a Vaidya metric matched to a flat radiation-dominated 
Universe, and in which the mass was found to grow proportional to the 
horizon 
size ($M(t)\to 0.06-0.08\, t$) at late times even for very small initial mass 
under the idealization of perfect radial inflow.

There are different approaches to estimating the accretion rate. Clearly it will 
be proportional to the surface area of the black hole, and to the energy density 
of the radiation background; the key question is the constant of 
proportionality, to which our results will prove extremely sensitive. The 
relevant length scales in the problem are the cosmological horizon size, the 
black hole radius, and potentially also the mean free path of the particles 
comprising the radiation background (which should be much less than the horizon 
size if the background is to be thermalized).

The two approaches available to estimate that constant are to use an absorption 
cross-section for radiation incident from a distance, or to consider the thermal 
properties of radiation near the event horizon. The former has the drawback that 
it assumes a long mean free path for the radiation (which cannot be true if the 
PBH radius is a significant fraction of the cosmological horizon), while in the 
latter the 
effect of the black hole geometry cannot readily be included. We will consider 
both and compare.

To use the cross-section of the black hole, one needs to take into account that 
the impact parameter for absorption of incoming radiation by the black hole is 
greater than the event horizon size, as the black hole bends the particle 
trajectory towards it. The black hole therefore has an effective radius $r_{{\rm 
eff}}$ for capturing particles. In the standard cosmology $r_{{\rm eff}}=3 
\sqrt{3}\, r_0/2$, where $r_0=2M/M_4^2$ is the Schwarzschild radius \cite{MTW}. 
In the case 
of an effectively five-dimensional black hole, the equivalent expression is 
$r_{{\rm eff}}=2 r_0$ \cite{Emp} where 
\begin{equation} 
r_0=\sqrt{\frac{8}{3 \pi}} \left(\frac{l}{l_4}\right)^{1/2} 
\left(\frac{M}{M_4}\right)^{1/2} l_4 \,. 
\end{equation}
The accretion rate is estimated by assuming first of all that the radiation is 
non-interacting, so that the radiation reaching the black hole at time $t$ has 
come from a known distance $x$. The fraction of the radiation starting at that 
point (assumed to have an isotropic momentum distribution) which is absorbed by 
the black hole is just the solid angle subtended by the black hole at that 
distance, and then adding up over all particles in a shell of width $dt$ at that 
distance yields
\begin{equation}
\label{calc1}
\frac{dM}{dt} = \pi r_{{\rm eff}}^2 \rho(t) \,,
\end{equation}
where the distance $x$ drops out of the calculation. This expression will 
continue to hold even in the presence of interactions, at least as long as the 
particle mean free path is much larger than the black hole radius, because in 
thermal equilibrium particles are as likely to scatter onto trajectories towards 
the black hole as off those trajectories. However it is unclear precisely what 
effect scattering near the black hole might give, particularly in regions where 
the black hole has significantly modified the geometry, and so we can expect 
this result to be modified by a factor of order unity. It is also unclear 
whether the depletion of the radiation in the vicinity of the black hole by the 
absorption might lead to a significant reduction in accretion efficiency. 
Finally, we note that this calculation is certainly na\"{\i}ve in ignoring 
spin and frequency dependent effects in the absorption cross-section 
\cite{Page,extra}.

A little more insight can be obtained by considering the thermal balance between 
the black hole and the background if they were at the same temperature. The 
black hole radiation leads to a mass loss given by the Stefan--Boltzmann law 
applied to the radius $r_{{\rm eff}}$ \cite{Emp}, except that it is also known 
that in the four-dimensional case the 
Stefan--Boltzmann law 
overestimates the emission by a factor 2.6 \cite{Page}, due to so-called 
grey-body factors allowing for the finite size of the black hole as seen by 
long-wavelength radiation (the majority of the power is emitted at wavelengths 
comparable to the event horizon radius) and the inclusion of spin-dependent 
effects. Such grey-body factors would also apply 
to absorption of long-wavelength radiation, and arguing that there should be 
balance when the cosmological and black hole temperature match suggests that in 
that situation Eq.~(\ref{calc1}) overestimates the absorption by a factor 
of around 2.6. However the regime of interest for accretion is when the 
background temperature is much larger than the black hole temperature, and then 
the grey-body factors should be much less important, supporting the 
normalization of Eq.~(\ref{calc1}). Were the cosmological 
temperature smaller the grey-body factors would be of increasing significance, 
but in that limit evaporation dominates in any case.

Care is however needed in applying thermal balance arguments in the braneworld 
case, because while accretion is taking place on the brane, the black hole can 
evaporate into both brane and bulk; in effect the black hole provides a route 
for leakage of energy from the brane via evaporation of gravitons. Even if the 
black hole is in thermal balance on the brane, it will not be in equilibrium 
with the bulk, and so would still lose mass. However the thermal balance 
argument can be used between the accretion and the evaporation onto the brane 
alone.

An alternative view is to consider the radiation background at a given time to 
extend smoothly all the way to the event horizon with an isotropic momentum 
distribution, and compute the flux entering the black hole. Consider a small 
shell of radius $dt$ at the event horizon, which will contain an energy density 
$4\pi r_0^2 \rho(t) dt$. However not all this radiation will be inwardly 
directed; the flux entering the black hole is one quarter of that energy density 
yielding
\begin{equation}
\label{calc2}
\frac{dM}{dt} = \pi r_0^2 \rho(t) \,.
\end{equation}
This is a factor of a few smaller than the estimate above. However this estimate 
is arguably more dubious. The approximation of an isotropic radiation 
distribution near the event horizon is unlikely to be good, as flux coming from 
the direction of the black hole will be absent, though one could argue that that 
this missing flux is correctly allowing for depletion of the radiation already 
absorbed by the black hole. More seriously, this estimate does not allow for the 
effect of the black hole geometry on the radiation, whereas the 
effective radius for absorption did in the previous calculation.

Summarizing the above, we can take the accretion rate to be
\begin{equation}
\label{accrate}
\frac{dM}{dt} = F \pi r_{{\rm eff}}^2 \rho(t) \,,
\end{equation}
where $F$ is a numerical constant measuring the accretion efficiency. It is 
conceivable that it might be as large as unity (particularly if the radiation 
mean free path is believed to be much larger than the event horizon radius), 
whereas other arguments suggest it may be somewhat smaller. However neither 
calculation is likely to be accurate if the black hole is a significant fraction 
of the horizon size, because then accretion is constrained by the amount of 
material actually accessible to the black hole. Unfortunately, it turns out that 
the results in the high-energy regime are exponentially sensitive to the value 
of $F$, and therefore we keep it in the calculations that follow. By contrast, 
the recent paper by Majumdar \cite{Maj} assumed $F=1$ throughout.

\section{Primordial black hole accretion}

\subsection{Accretion in the standard cosmology}

As a warm up, we can apply the above formalism in the standard cosmology, where 
however it will turn out that the approximations made are unlikely to be valid.

Suppose a PBH is formed at a time $t_{{\rm i}}$ in the early phases of a
radiation-dominated Universe, with mass 
\begin{equation} 
M_{{\rm i}}=f M_H(t_{{\rm i}})=f M_4^2 t_{{\rm i}} \,. 
\end{equation}
Here $M_H(t_{{\rm i}})$ is the horizon mass and $f<1$ denotes what fraction of 
the horizon mass the black hole comprises. Using Eq.~(\ref{accrate}) with the 
four-dimensional 
Schwarzschild solution and the radiation energy density
\begin{equation}
\rho = \frac{3M_4^2}{32\pi t^2} \,,
\end{equation}
gives
\begin{equation} 
\frac{dM}{dt}=\frac{81}{32} F M_4^{-2} \frac{M^2}{t^2} \,. 
\end{equation} 
Integrating from a time $t_{{\rm i}}$ when the black hole mass is $M_{{\rm i}}$ 
leads to
\begin{equation} 
\label{QS-stand}
M(t)=M_{{\rm i}} \left[ \frac{t/t_{{\rm i}}}{A \, t/t_{{\rm i}} + (1-A)} \right] 
\quad ; \quad A\equiv 1-\frac{81}{32} Ff \,. 
\end{equation}

The behaviour of $M(t)$ crucially depends on the sign and magnitude of
the factor $A$. If it is positive, the black hole mass asymptotes to
\begin{equation} 
M_{\infty}=\frac{M_{{\rm i}}}{A} \,. 
\end{equation}
If $F$ and/or $f$ are small (small efficiency or initial PBH size), the
factor will be close to 1, and the asymptotic
mass will not be much bigger than the initial mass. But if $A$
is small ($Ff\to 0.4$), the black hole mass will grow nearly
as fast as the horizon mass, until $t/t_{{\rm i}}\approx 1/A$ or until
the radiation-dominated regime comes to an end. The factor $A$ can
even be negative for $Ff>0.4$, in which case Eq.~(\ref{QS-stand}) gives
a diverging mass in a finite time, violating causality and thus certainly 
indicating that the approximations used have broken down. 

However, in order to obtain a final mass that is much bigger than the
initial mass, it is required that the initial PBH radius must be close to 74 
percent ($=0.4^{1/3}$) of the horizon radius (for efficiency $F=1$), or more
(for $F<1$). For such large PBHs use of the quasi-static approximation (where 
the black hole is represented as a series of Schwarzschild solutions as the mass 
increases), becomes
questionable. But in the case of PBHs the initial size of the black
hole $\it{must}$ be of the order of the horizon, to overcome the fluid
pressure. Carr and Hawking \cite{CarrHawk} proved there was no black
hole solution embedded in a Friedmann background in which the black
hole could grow at the same rate as the cosmic horizon. They then concluded
that the black hole could only grow less fast than the horizon. After a short
period of time its size $\it{would}$ have become much smaller than the
horizon, at which point the quasi-static approximation could be used to confirm 
no further accretion. This 
result has become widely accepted, though see Refs.~\cite{Hacyan,accrete}.

\subsection{Accretion in the high-energy regime of the braneworld scenario}

We now turn to the high-energy regime of the braneworld scenario. If black holes 
are to behave as five-dimensional ones, we showed \cite{GCL} that they must form 
during the high-energy regime, and that it was possible for such PBHs to survive 
even to the present. In order to interpret possible observational signatures, we 
therefore need to understand accretion in the high-energy regime. In this 
section we consider only accretion, and in the following section we will 
consider the combined effects of accretion and evaporation.

If a black hole forms at a time $t_{{\rm i}}$ in the high-energy regime of the
Randall--Sundrum Type II scenario, its size will necessarily be smaller
than the AdS radius $l$, and hence described as a five-dimensional
Schwarzschild black hole. We take its initial mass to be a fraction $f$ of the 
horizon mass (now computed using formulae relevant to the high-energy regime 
\cite{GCL})
\begin{equation} 
M_{{\rm i}}=f\, M_H(t_{{\rm i}})=16\,f M_4 
\left(\frac{l}{l_4}\right)^{-1}\left(\frac{t_{{\rm i}}}{t_4}\right)^2\,.
\end{equation}
The growth rate is again given by Eq.~(\ref{accrate}), where we must now
employ the energy density in the high-energy phase 
\begin{equation} 
\rho=\frac{3}{32 \pi}\frac{M_4^2}{t_{{\rm c}}\,t}\,, 
\end{equation}
with the transition time between high-energy and standard regimes
given by $t_{{\rm c}}=l/2$, and use the five-dimensional expressions for the 
event 
horizon radius and capture cross-section. This gives
\begin{equation}
\frac{dM}{dt} = \frac{2F}{\pi} \, \frac{M}{t} \,.
\end{equation}

Integrating from the time of formation onwards results in 
\begin{equation} 
\label{justaccmass}
M(t)= M_{{\rm i}}\,\left(\frac{t}{t_{{\rm i}}}\right)^{2 F/\pi}\,. 
\end{equation} 
In this case we see that the black hole mass always grows
less fast than the horizon mass $M_H(t)\propto t^2$ (provided $F$ is not 
significantly greater than unity), in contrast to the situation in the
standard cosmology. Accordingly, the approximations made in the calculation 
become more reliable as time goes by. Our result agrees with that of Majumdar 
\cite{Maj}, who however assumed $F =1$ throughout.

This calculation shows that, regardless of the initial mass, black holes can 
experience significant growth, with the power-law depending sensitively on the 
accretion efficiency $F$. The uncertainty in its precise value lends 
considerable uncertainty to the accretion growth of the PBHs during the 
high-energy phase.

Once the high-energy phase ends, the cosmological background evolution becomes 
the standard one, but the PBH remains in the five-dimensional regime. To 
determine whether accretion still continues, we now need to apply 
Eq.~(\ref{accrate}) with the five-dimensional PBH properties but the standard 
cosmological evolution, taking as initial condition the mass of the PBH at the 
transition time $t_{{\rm c}}$ and bearing in mind that the black hole will be 
much smaller than the horizon size at this point. Using these equations gives
\begin{equation} 
M(t)=M(t_{{\rm c}})\, {\rm exp}\left[\frac{2F}{\pi}\left( 1-\frac{t_{{\rm 
c}}}{t}\right) \right]\,. 
\end{equation}
indicating a further growth by only a factor of order unity (comparable to the 
expected uncertainties in the calculation).

We therefore conclude that accretion swiftly becomes unimportant as the standard 
cosmology sets in for five-dimensional black holes as well as four-dimensional 
ones. Accordingly, it is the slower decrease of the background density during 
the high-energy regime which makes accretion important, rather than the change 
in black hole properties.

\section{Combining accretion and evaporation}

We now take both accretion and evaporation into account in the
high-energy regime\footnote{Some work loosely related to this, but concerning 
astrophysical accretion by collisionally-produced mini black holes in a TeV 
gravity model, can be found in Ref.~\cite{TeV}.}
\begin{equation} 
\frac{dM}{dt}=\left(\frac{dM}{dt}\right)_{{\rm acc}}+\left(\frac{dM}{dt}\right)_
{{\rm evap}} \label{acc+evap} \,,
\end{equation}
where
\begin{equation} 
\left(\frac{dM}{dt}\right)_{{\rm acc}}= \frac{q}{2}\,\frac{M}{t} \,,
\end{equation}
and
\begin{equation} 
\left(\frac{dM}{dt}\right)_{{\rm evap}}=-\frac{\tilde g\, M_5^3}{2 M} \,.
\end{equation}
(see Ref.~\cite{GCL}), where for convenience we have defined
\begin{equation} 
q\equiv \frac{4 F}{\pi}\,.
\end{equation}
Assuming the four-dimensional cosmological constant 
vanishes, the relation between the five-dimensional fundamental mass scale and
the Planck mass is given by
\begin{equation} 
M_5=M_4 \left(\frac{l}{l_4}\right)^{-1/3}\,.
\end{equation}

In the above expression we have defined
\begin{equation}
\tilde g\approx \frac{0.0062}{G_{{\rm brane}}} \;g_{{\rm brane}} 
+\frac{0.0031}{G_{{\rm bulk}}} \;g_{{\rm bulk}} \,.
\end{equation}
where $g_{{\rm brane}}$ is the usual number of degrees of freedom into
which the black hole can evaporate, while 
$g_{{\rm bulk}}= \mathcal{O}(1)$ is the number of bulk degrees of freedom 
(in the simplest case just the five polarization states of the
graviton). For the most part, the black hole's energy is lost through Hawking
radiation on the brane. 
As an example we mention the case where the black 
hole emits only massless particles, for which $g_{\rm{brane}}=7.25$ and 
$\tilde g=0.023$. Unlike in Ref.~\cite{GCL}, we have written the grey-body 
factors $G_{{\rm brane}}$ and $G_{{\rm bulk}}$ explicitly; in the standard 
cosmology the grey-body factor is equal to 2.6, but precise values are not known 
for the braneworld. For thermal balance between the accretion and the 
evaporation onto the brane, the accretion efficiency $F$ should equal $1/G_{{\rm 
brane}}$ when the temperatures are equal; however as discussed in Section~II the 
absorption grey-body factors should approach one if the background temperature 
is much greater than the black hole temperature.

In the standard cosmology Eq.~(\ref{acc+evap}) does not have an
analytical solution, but it does in the present case: 
\begin{eqnarray} 
M(t)/M_{{\rm i}} & = &  \label{mass}\\
 && \hspace*{-40pt} \left\lbrace\left(\frac{t}{t_{{\rm i}}}\right)^q -
 	\frac{\tilde g}{4 \sqrt{f}}\,\frac{1}{1-q}
 	\frac{M_5^{3/2}}{M_{{\rm i}}^{3/2}} \left[ \left( 
 	\frac{t}{t_{{\rm i}}}\right)-\left(\frac{t}{t_{{\rm i}}}
 	\right)^q \right] \right\rbrace^{1/2} \,.  \nonumber 
\end{eqnarray}
An expression of this form was found by Majumdar \cite{Maj}, though again for 
the specific case of efficiency $F=1$ (i.e.~$q=4/\pi$). However, a very slight 
change in the efficiency can change the qualitative behaviour.
If $q>1$ (efficiency $F>0.78$) the PBH will steadily grow
until the cosmological transition time $t_{{\rm c}}$ is reached (provided 
$M_{{\rm i}} > M_5$). If $q<1$ 
(efficiency $F<0.78$), the mass
loss term will grow faster than the mass gain term. However, assuming
that the initial black hole mass is large compared to the fundamental mass 
scale, $M_{{\rm i}}\gg M_5$, the loss term will initially be much smaller
than the gain term. In other words, if a black hole forms with a mass
of order the horizon mass at a time $t_{{\rm i}}\gg t_5$, then the initial black
hole temperature will be much lower than the temperature of the
radiation background and evaporation can initially be neglected. 

For many choices of parameters, evaporation is then negligible throughout the 
high-energy regime. But
for sufficiently low accretion efficiency, the mass growth can come to
a halt whilst the black hole is still in the high-energy regime.\footnote{In the 
limit $q 
\to 1$ the mass growth decreases
logarithmically, and if $M_{{\rm i}}\gg M_5$, evaporation can be neglected until 
the 
standard regime is reached.}
This `halt' time is obtained from $dM/dt=0$ and reads
\begin{equation} 
\left(\frac{t_{{\rm h}}}{t_{{\rm i}}}\right)^{1-q}=q\left[1+(1-q)\,\frac{4 
\sqrt{f}}
{\tilde g} \left(\frac{M_{{\rm i}}}{M_5}\right)^{3/2}\right] \,.
\end{equation}  
If $M_{{\rm i}}\gg M_5$ we can neglect the first term, to obtain
\begin{equation} 
\left(\frac{t_{{\rm h}}}{t_{{\rm i}}}\right)^{1-q}\approx q\,(1-q)\,\frac{4 
\sqrt{f}}
{\tilde g} \left(\frac{M_{{\rm i}}}{M_5}\right)^{3/2} \,.
\end{equation}  
In order for the derivation to be valid the halt time must satisfy $t_{{\rm 
h}}<t_{{\rm c}}$, or equivalently
\begin{equation} 
\left(\frac{t_{{\rm i}}}{t_4}\right)^{4-q}< \frac{\tilde 
g}{f^2}\,\frac{2^{q-9}}{q (1-q)} 
\left(\frac{l}{l_4}\right)^{2-q}\,. \label{thaltcondition}
\end{equation}
The above condition will be satisfied if $l$ is large enough (a long high-energy 
regime)
and/or if $M_{{\rm i}}$ is small enough (a high initial PBH temperature),
depending on the value of $q$. Put another way,
for a given $l$ and $M_{{\rm i}}$ there is a minimum accretion efficiency which 
ensures that
neglecting the mass loss through evaporation is justified all the way
up to $t=t_{{\rm c}}$.

As an example we bear in mind that the AdS radius $l$ is constrained by
experiment as
\begin{equation} 
l<10^{31} l_4\,.
\end{equation}
If there was a period of high-energy inflation, the requirement that
gravitational waves do not lead to excessive anisotropies in the CMB
leads to a lower limit on the horizon mass (and hence on the PBH mass)
at the end of inflation, namely \cite{GCL}
\begin{equation} 
M_{{\rm i}}>2\times 10^6 M_5\,.
\end{equation}
For the extreme values of $M_{{\rm i}}$
and $l$ (and hence for all other values), we find that an efficiency better than 
$30
\%$ ($q>0.39$) is sufficient to ensure mass growth all the way up to the cosmic
transition time.

We now ask what is the total lifetime $t_{\rm{evap}}$ of a black hole
that reaches a halt time in the high-energy regime. Provided we can trust
Eq.~(\ref{mass}) until total evaporation $M(t_{{\rm evap}})=0$, we find     
\begin{equation} 
t_{{\rm evap}}=q^{1/(q-1)} t_{{\rm h}}\,.
\end{equation}
Unless $q$ is very close to zero, we see that the total lifetime will be
of the same order as the halt time (taking into account the
possibility that the final stage of the black hole's lifetime is in the
standard regime only $\it{decreases}$ the estimate for $t_{{\rm evap}}$, as
it turns off the accretion term). Such black holes evaporate long before any 
observational constraints can be brought to bear, and we therefore conclude that 
PBHs in the regime where accretion in the high-energy regime can halt are not of 
interest.

To summarize, for a PBH with a lifetime $t_{{\rm evap}}\gg
t_{{\rm c}}$ we need consider only two situations in analyzing their evolution 
forwards 
in time. For black holes forming after the end of the high-energy regime,
accretion is never important. For those forming within the high-energy regime, 
it is always a good approximation to neglect evaporation up until the cosmic 
transition time, after which accretion can be ignored and evaporation dominates. 

\section{Conclusions}

We have explored the possibility that PBHs might accrete from the cosmological 
background in the case of braneworld cosmology, and found, in agreement with a 
recent paper by Majumdar \cite{Maj}, that significant growth is possible. 
However, we have highlighted the extreme sensitivity of the resulting growth to 
the assumed accretion efficiency, which cannot be accurately computed. Accretion 
therefore adds considerable uncertainty to the evolution of individual PBHs 
after formation in the braneworld scenario.

Since accretion ends once the standard cosmology is restored (whether the 
PBHs are effectively four-dimensional or five-dimensional), we stress that 
accretion does not have any implications for interpreting observations in terms 
of the PBH density {\em at evaporation}. It does however impact on how those 
constraints are interpretted in terms of formation rates in the early Universe.
We will be providing a detailed analysis of observational constraints on 
braneworld PBHs in a forthcoming paper (Clancy, Guedens and Liddle).

\begin{acknowledgments}
D.C.~was supported by PPARC and A.R.L.~in part by the Leverhulme Trust. 
R.G.~would like to thank John Barrow and Malcolm Fairbairn for discussions.
\end{acknowledgments}


\end{document}